**Productive self/vulnerable body: self-tracking, overworking culture, and conflicted data practices**

**Elise Li Zheng**

**Abstract**

Self-tracking has gathered scholarly attention as digital devices and wearables gain popularity. The collection, analysis, and interpretation of personal data signifies an individualized way of health governance as people are demanded to build a responsible self by internalizing norms and changing unhealthy behaviors. However, the technological promises often bear conflicts with various social factors such as a strenuous schedule, a lack of motivation, stress, and anxieties, which fail to deliver health outcomes. To re-problematize the phenomenon, this paper situates self-tracking in an overworking culture in China and draws on semi-structured and in-depth interviews with overworking individuals to reveal the patterns in users' interactions and interpretations with self-tracking data. It builds on the current literature of self-tracking and engages with theories from Science and Technology Studies (STS), especially sociomaterial assemblages (Lupton 2016) and technological mediation (Verbeek 2005), to study self-tracking in a contextualized way which connects the micro (data reading, visualization, and affective elements in design) with the macro (work and workplaces, socioeconomic and political background) contexts of self-tracking. Drawing on investigation of the social context that users of self-tracking technologies internalize, reflect, or resist, the paper argues that the productivity- and value-oriented assumptions and workplace culture shape the imaginary of intensive (and sometimes impossible) self-care and health, an "involution" of competence embedded in the technological design and users' affective experiences. Users respond by enacting different design elements and social contexts to frame two distinctive data practices of self-tracking: to perform, or to calibrate. Users enact such data practices to navigate stress and anxieties under an overworking culture and social and economic pressure and to negotiate a selfhood by pursuing a "status-being" which encompasses personal identity, physical performance, and mental wellbeing. This paper dissects various pathways that technological designs and production processes affect people's health narratives and calls for novel approaches to "thicken" the health data that take consideration of people's living experiences and social context.

**Introduction**

"My colleagues are too afraid of going home too early after work." Once a junior software engineer, Mr. Zhuge started an online campaign called "996.icu" on GitHub.com, protesting the grueling long working hours in China's tech sector which Alibaba's CEO Jack Ma called a "blessing" for the workforce (Lin & Zhong, 2019). "996" means "9 a.m. to 6 p.m., 6 days per week," an overworking scheme that has been popularized by tech companies in China and become widespread across the private sector. "ICU," the intensive care unit, indicates the dire health risks of overworking, and the news about early death (usually results of undetected cardiac arrest) is not rare (Wang, 2021).

The overworking culture in China have sparked fierce discussions both in China and around the globe, and the dire health situations of the workforce, not limited to the tech sector, became one of the greatest concerns. For the Chinese workforce – increasingly individualized and labeled as new "middle-class" with potential consumption power – overworking has been intimately tied to the need for self-care and nurtured a fast-growing market of health-tech, such as smart watches, wristbands, and self-tracking applications (apps), targeting the burnout by promising a healthier body and a well-managed life.

The freedom, and sometimes overabundance, of choices of self-care speaks to the neoliberal approaches of health in post-socialist China. However, in the context of an overworking culture, as I show in this paper, self-care becomes ambivalent and operates on multiple registers, and sometimes in conflicts with each other. The bodies – imagined and built through self-tracking technologies – are not just strong, healthy, and capable in a competitive workplace, but also fragile and at-risk, faced with stress, exploitations, and burnout that needs protection from work. This paper, therefore, seeks to unveil such ambivalence and multiplicity of self-care, showing the entanglement of the technological and social process in the framing of health-related, data practices. At the end, this paper shows the limitations of current digital health data measurements and models, and calls for alternative, "thick" understandings of health data that better reflect the social and political factors and living experiences with technologies.

## Situating Self-care in Current China: througha Looking Glass to an Overworking Self

The Japanese word *karōshi* (過労死) emerged a few decades ago might hint at the common root of overworking shared by East Asian countries, yet the new trends of "996" in China encompasses not only organizational control of the workforce, but also the tension between individual livings and societal aspirations. The imaginary of "high quality" (*gao suzhi*, 高素质), hardworking citizens encompasses patriotic, future-oriented sentiments entangled with value defined by an authoritarian government (Kipnis 2006, So 2007). Especially after the Xi authority coined the idea of "Chinese Dream," the incorporation of individual approach and nationalist orientation, requires not only the spirit of hardworking and dedication, but also the sense of productivity, self- entrepreneurship and responsibility (Gow, 2017; Hann, 2022; Harvey, 2005; Lindtner, 2020; Rofel, 2007; Yan, 2010). However, faced with economic slowdown and uncertainties, the creed of progress turns into a society-wide tension described as "involution"

(*neijuan*, 内卷), a term borrowed from historical research of Asian agrarian economy[1] to negatively denote a sense of never-ending, sometimes pointless competitions for growth.

Not only work but also self-care falls into the endless cycle of "involution" for improvement – the

building of a healthy, responsible, and well-informed individual. The *Health China 2030 Strategic Plan*

*Outline* released by the Chinese State Council (Xinhua Agency, 2020) prioritized the promotion of "self-

initiated and self-discipline behaviors." It is exemplified in the high requirement of health self-management, especially among the mid-30s and 40s, who are burdened with various health risks due to stress and burnout (Dingxiang Doctor, 2020; J. Li, 2016; T. Yang et al., 2012). Self-tracking devices have gained popularity, making it possible to track bodily metrics and behaviors in well-designed schemes of self-management, alongside vast scales of collection of personal data. It speaks to the orientations of neoliberal health, in which individuals bear responsibilities to manage uncertainties and risks (Beck,

1992; Beck et al., 1994; Lupton, 1995). Health has transformed into a personal choice that someone can

make (Brown & Webster, 2004). The seamless, real-time and intimate interaction between bodies and

technologies creates a culture of self-surveillance (Ajana, 2017, 2018; Lupton, 2016), in which the

collection of bodily biometrics and self-regulation is seen as responsible biological citizens' agency

(Foucault, 1990, 1991; Rabinow & Rose, 2006; Rose, 1999, 2007; Sanders, 2017; Schüll, 2016).

However, the building of selfhood with science and technology in China, especially at the intersection of

work and life, carries important caveats – the ostensible individual "choices" that the market promises

often collides with traditional, collective, and institutional powers. Research in psychiatry (Borovoy &

Zhang, 2017; J. Yang, 2017; Zhang, 2017, 2018), environmental health (Kohrman, 2020; Liu, 2021) and maternity (J. Li, 2021) have shown that far from adopting a universal approach, the practices of knowledge, healthcare and technology are deeply intertwined with local context, and the idea of self-care produces their identity and capacity for fulfilling certain social and political roles and shows one's sense of power and privileges (Zhang & Ong, 2015). The precarity and uncertainty of work (Moore, 2018) and inevitability of "involution" in a slowed down economy (X. Li, 2022) also highlight the conflicts of self-care in China – is the self-care aiming at building a self and body through individualized productivity, or to protect it from all the obligations? In the next section I investigate the practices of self-care among the Chinese workforce based on my fieldwork.

---

[1] "Involution" (*neijuan*, 内卷) was first used by historian Clifford Geertz to describe rice-cultivation in Indonesia
with greater social complexity but lack of significant technological or political change, a "change without progress"
(Geertz, 1963 & 1991). It has become a buzzword in China in the beginning of 2020s (Liu 2021), reflecting a certain
type of social anxiety among younger generations as the economic growth slows down and competition for
resources and upward mobility becomes ever fierce. It is borrowed and used in numerous situations without a clear
scholarly definition (Wang 2022), yet Chinese sociologists such as Biao Xiang have talked about it on mass media
as a cultural phenomenon (Wang and Ge, 2022).

## Methods and Materials: examining ambivalence and multiplicity

Self-tracking is not only a manifestation of self-discipline and awareness of health responsibilities, but also a series of data practices that encompasses the process of knowing, doing, and reflecting each individual user's bodily status and living experiences (Fors et al., 2020; Pantzar & Ruckenstein, 2017). One important task is to connect micro (personal, affective, behavioral) with macro (social, cultural, political). STS scholars have developed theoretical and methodological frameworks that dissect such sociotechnical entanglements. The idea of "technological mediation" (Ihde, 1990; Rosenberger & Verbeek, 2015; Verbeek, 2006) adopts a phenomenological approach that looks at the interfaces and interactions between human and technologies and sees technologies as connecting points of users and their outside world. The embodiment of actions and interpretation of meanings highlight the processes and focus on the building and cementing of human-technology relationship, and a new form of bodily technique (Abend & Fuchs, 2016; Mauss, 1973) – where the mundane knowing and doing of human body is given social and cultural significance. From there, the use of technologies becomes *narrated* and *framed* experiences, and the phenomena transfer into *practices*. The study of practices accentuates the situated and performative nature of technological encounters (Suchman, 2007), displaying complexities and contingencies that address multiplicity and differences (Berg & Mol, 1998; Mol, 2003, 2008; Mol &

Berg, 1994). In addition, it helps the technological inquiries venture into the the realm of power and resources during conflict and resistances (Clarke & Montini, 1993), and renders a critical view on the distribution of agency, control, and capacity among human and nonhuman in different contexts of use (Fotopoulou, 2019; Lupton & Smith, 2018; Marent & Henwood, 2021; Pols et al., 2019; Schüll, 2012).

Anchored in the studying of technological practices, this study uses first-hand, empirical materials that I gathered during fieldworks between 2019 and 2022. I have conducted 94 semi-structured interviews and 15 follow-up, in-depth interviews with self-trackers. The participants were recruited online through WeChat (the biggest social media/chat app in China). The open recruitment messages were posted in WeChat Moment and chat groups, and snowball sampled throughout participants networks. The selection criteria are those who live in urban areas in China, self-report experiencing overworking, stress and/or burnout in workplaces, and have regularly used at least one type of self-tracking device/application to monitor health. About half of them self-reported in an overworking situation on a regular basis (approx. weekly work hours>50). The sampling method ensures the enrolment of individuals who are likely to meet the selection criteria and their experiences triangulate each other, such as people work in the same company. It bears limitations of representations and lacks diversity in socioeconomic status, which should be addressed in future research.

*Table 1 interviewees' background*

| Gender | | Age | | Job sector[2] | | Job title | | Overworking situation | |
|---|---|---|---|---|---|---|---|---|---|
| Male | 38 | 18-29 | 37 | Tech sector | 53 | Engineer | 16 | On a regular basis | 46 |
| Female | 56 | 30-40 | 49 | Other | 41 | Product | 14 | Occasional overwork | 23 |
| | | >41 | 8 | | | Market | 15 | No overwork, but stress | 25 |
| | | | | | | Content | 11 | | |
| | | | | | | Consultant | 8 | | |

[2] There is a certain extent of vagueness when someone describe their job placement. Here, tech sector includes companies who makes revenue by selling technological/internet devices, services, or digital contents; the workforce is commonly described as "knowledge workers," distinct from manufacture or service sector jobs. The job titles are typically software engineers/developers, designers, product managers, marketers, content managers, among others. The "996" scheme was firstly popularized by big-tech companies in China such as Huawei and Alibaba, and the tech sector work in China is characterized as growth-oriented, highly competitive, self-initiative and with potential of upward mobility. However, such overworking culture is not exclusive to the tech sector – it has long been a common practice among jobs such as physicians and consultants. See X. Li 2022.

| | |
|---|---|
| Physician | 2 |
| Other | 28 |

All the interviews were conducted online using WeChat video, and audio recordings were transcribed and coded in nVivo for qualitative, thematic analysis. The questions asked about their daily routines, workloads, self-tracking behaviors, data interpretation, and health perceptions. The first round of coding was focused on the theme and topics of life routines, workplaces, and self-tracking data; the second round was to draw connections to show the composition of technological practices and how they are situated in their work and life settings. The cross-coding reveals how meanings are produced and communicated, how materials interplay with human agency, how skills/competence are built, and how habits and even conventions are formed and performed (Ingram et al., 2007; Shove et al., 2012). The language is Chinese, and all the quotes in this paper is translated verbatim. Aside from the interviews, other materials such as news releases from wearable and app makers, news reports, and design features of the apps and devices are critically examined to triangulate users' experiences. Drawing on the framework of technological practices, I ask:

1) How is the pursuit of productivity, as well as the affective experience of "involution," enacted during self-tracking?

2) How are multiplicities and differences performed flexibly through engagement with personal data?

3) What underlies the differences, alternatives, and even resistance against the formative way of self-surveillance, and how are they connected to the current the political economy?

**Findings: productivity with external validation**

*Meanings, skills, and involution of self-care*

One critical component of self-tracking is the collecting and reading of personal data – notably biometric and behavioral data collected by the sensors, and sometimes recorded through users' active engagements such as journaling, logging, checking-in, and sharing. Certain visualizations and interactive features helped users cultivate a set of meanings, skills, and habits, which interplay with the material designs, to perform self-care effectively and efficiently, sometimes at the expense of reducing body and living experiences into abstract numbers that can be managed and governed (Ruckenstein & Pantzar, 2017; Ruffino, 2017; Smith & Vonthethoff, 2017).

For example, the reward-giving designs are positively received across self-tracking apps and devices. Jia, who works for an e-commerce company, was trying to finish a "Perfect Month Challenge" on her Apple Watch. That means "closing the rings" every day for a month – achieving goals for standing up in all 12 hours, burning over 400kcal calories, and exercising 20 minutes. She was full-on, until Shanghai hit a lockdown due to COVID-19 outbreak, and she suddenly lost the streak. She showed me that one of her friends was also on this challenge and posted in her social media timeline – the calendar with all closed rings, and a funny comment, "this 'digital cuff' – it works."

This type of so-called "persuasive technologies" (Fogg, 2002) is often made to embody the technological experiences to incentivize continuous use and engagements. The practices are centered around personal plans and goals to stimulate a sense of achievement, especially when performance screenshots are captured right after exercise sessions or when users scroll back to summarize a certain period of activities. Users perform these behavioral data as external validation of voluntary hard work – and sometimes with a sense of self-deprecative humor, that the technology acts as a source of robotic authority to give validation. Users deployed the phrase "refresh" (*shua*, 刷, originally means the F5 refresh key on the keyboard) to describe their actions, indicating getting credits by repeating a fixed action and getting external validation. In practice, users try to make up steps and exercise time on their wearables by turning on the exercise record when running errands on foot or on bikes, turning mundane actions into a productive way to accumulate bodily work.

> *There is an impulse to 'refresh' my steps (when I have my wristband on) …I would walk out for lunch on purpose. (#2-19, male, uses Apple Watch)*

> *I will try to "reach" the goal when it is close. For example, for short- or mid-distance commute, I'll use bike and turn on the record to 'refresh' some numbers. (#2-16, male, uses Apple Watch)*

> *(after seeing social media posts) I bought the smart rope-jumping thing to refresh some numbers. 3000+(counts) means good fat-burning. It is definitely faster than running. (#1-20, uses a smart rope-jumping device)*

Some apps encourage users to "clock-in" (*daka*, 打卡) daily, shown as crossing off certain tasks[3].

> *I will clock-in every time I finish an exercise, giving myself some achievement, posting on my timeline and stuff…I feel great as if I have finished something great. (#1-2, female, uses Keep)*

> *I really need some encouragement. I'll tick off things on "clock-in" apps. It goes into my monthly summary, that I have exercised a lot…I have clocked-in nearly 1000 times. (#2-23, female, uses Keep)*

Apple Watch gives daily "suggestion" of calorie consumption and frequently adjust goals according to the user's daily physical activities – usually pushes towards higher goals with a persistent idea of improvement. Fitness apps, such as *Keep* in China which carries plan-making functions tailored to losing weight or gaining muscles, pan out linear plans of progress. This speaks to the assumption that as part of datafication, all activities can be broken down into routine, standard, sequential tasks and could be self-automated with the help of technology (Pharabod et al., 2013; Wajcman, 2019). Etkin et al. (2016) argue against gamification elements since such design transforms enjoyment into the language of work and damages long-term outcomes. But according to the users, the quantified rewards don't replace users' own willingness and motivations. Quite the contrary, the meaning of productivity – non-stopping growth and improvement – is produced and internalized through self-tracking records with work/productivity as both the means and the ends. The productivity-oriented bodily data necessitates the willingness of change as forms of self-care.

---

[3] The term "clock-in" (*daka*, 打卡; sometimes referred as "check-in") is a term originally used to describe workers using a punch card to record their presence. It has been appropriated by technological designs to indicate the validation of having something done.

> *The data shows all the details, but I have to have a willingness to move forward, and the goal that I really want to accomplish. (#2-35, uses Apple Watch)*

However, these automated, "convenient" (#1-12, #2-3, #2-16) and "streamlined" (#2-51) settings demand stringent self-regulation and planning, especially to those who have a tight schedule or work long hours. Fitness goals and the meaning of a hardworking, over-achieving self override the meaning of health, as one interviewee (#2-13) associated "more of self-discipline rather than health eating" with her month-long eating plan and tracking records. Users will have a honeymoon with the sense of rewards, yet the extra pressure to keep up is translated into languages of "involution," the same sentiment against endless growth, goal-chasing, and competition. Ran, in his 40s and works as a manager for a supply chain company, purchased a fitness plan from *Keep*. His personal data from a smart scale was also synchronized as well as tracking daily meal intakes. The exercise routine went on a "level-up" mode according to "progress" of his bodily data, until he injured himself during an intensive HIIT training session. He carried on for about a year, but recently ditched the plan. He mentioned work-related stress, tiredness, and lacking willingness, "just don't want to move (after work)." But there is always a sentiment of self-blame, as he said:

> *I installed the app so that I can exercise any time… (I worked) until 11 p.m. and realized that today's exercise plan hasn't been carried out…there's a frustration that why someone else can do but I can't?*

Others expressed similar sentiments that health become demanding in the language of productivity:

*I bought a watch for my workout…I don't have a coach or trainer anymore, everything must be controlled by myself…the initial feedback (on data) was very good, it gave me a lot of incentives. I walked 3km to and back from the gym every day, even after work around 8pm…then I questioned myself, do I have to be that strict? No social, my life is just work, gym, and home. (#2-54, female, in her 30s, works in a state-owned enterprise)*

*Around the year of 2018 I moved to Beijing, and had a period of unstable time. I didn't have time to think about (the plans) …I needed strong self-discipline to open (the app). It is stressful to stick to a higher standard of health." (#1-1, female, was on a tight diet while working in digital media)*

Such sentiment with technological design is not only experienced through designs and interactive features, but also shaped by the ecosystems of commercial self-tracking products and the physical/social environment around overworking individuals.

*Invisible work in health and workplaces*

The productivity-oriented design is embedded in the making of the health-tech "ecosystem," where users act as well-informed consumers who choose to improve health performance through consumption (Hardey, 2019). The intensity originates from the emerging field that attracts venture capital (VC) with the promise of a potential market and scalable business model. App makers and service providers in this allegedly fast-growing market, in fact, have not lived up to a profit-winning business[4]. The promise of future growth has intensified self-tracking – the need to expand scales, accumulate users, and boost engagements for values from users' "prosumption" (Millington 2016). The self-tracking practices are imagined as a desired path of progress from novices to well-trained, well-informed consumers of online courses, plan subscriptions, offline gyms, fitness attires and healthy food packages. The technological designs thus are geared towards incentivizing active using and habituation, to an extent that some users complained it being "heavy" ("function creep").

---

[4] In China, the company which makes the most popular fitness app *Keep* went public in January 2022 without current profit but a proposed ecosystem in their IPO.

The habituation of health-related self-tracking practices is incorporated into the busy workplaces. As work becomes more individualized and precarious, productivity becomes a measurement of both products and aspects of self and bodily states (Moore, 2018; Wajcman, 2019). In the users' narratives, the boundaries between personal life and work have been blurred, and individuals must take extra responsibilities to "manage" the intricate relationship between personal spaces and workplaces. Interviewees associated such practices of "arrangements" when talked about how they integrate their self-tracking practices such as steps, exercising plans, and sleeping patterns into work and life, rather than going through smooth, smart interactions with devices.

This type of managing is more commonly seen in the tech sector, and the need is more imperative among those who work from home, or during COVID-19 lockdowns. The popular notion of work-life balance from Silicon Valley is adopted by Chinese tech companies by installing workplace gyms, massage rooms, and fitness meal services as part of wellness programs (*fuli*, 福利) in parallel with monetary bonuses. It weirdly fits within an overworking culture – employees enjoy "freedom" to arrange their workout by making time during lunch break, after dinner before overtime work, in the late night or early morning, to utilize the facilities and services. An interviewee working in a medium-sized tech company said that she often incorporates fitness app's exercise into her daily schedule (#1-13). Two interviewees (#2-14, #1-3) told that they "block off" a certain time, so that they could be shielded from work messages and meetings while going for arranged exercises. One interviewee (#2-63) frequently uses gym hours to fulfil his "996" hours when he must stay around the office but doesn't get too much work at hand. Such scenarios are reinforced by the use of technologies, as they quantify, visualize, and materialize individual efforts, creating the intricately overlapping space and time as an "ubiquitous technoscope" (Wajcman, 2015) that normalizes multitasking. Workplaces encourage sharing fitness records in "clock-in groups (*daka qun*, 打

卡群)", in which self-tracking records are proof of performances. Aside from gym workouts, users respond to notifications to alert sedentary seating, perform micro exercises during meetings, and accumulate step goals during lunch breaks (as interviewee #2-82 said, "standing up once in a while and close the standing ring is the easiest of all.").

The mental and physical resources are required yet in short supply in an overworking environment. As an interviewee complained, the fatigue after work "sink myself into my couch" and her mental resource is "only sufficient for no-brain entertainment" (#2-58). Therefore, people have developed different coping strategies and/or alternative narratives around self-tracking data. The multiple practices and flexible interpretations are associated with, and sometimes enacted by, the productivity-oriented data and technological design, but acquire various meanings and skillsets with underlying needs for control and balance during events of stress and uncertainty.

*Performance or "status being:" enacting bodily metrics*

Pei, in his early 30s, works for a small tech company who just underwent a period of typical overwork at a new position. He only slept 4-5 hours each night in the past week but still tried to maintain his exercise habit amid a busy schedule. He didn't know when the work would conclude each day, yet he still booked some fitness classes and "arrange other things around them." He chose to attend classes at a lower frequency but of more intensive ones. He recalled the most recent exercise by looking at his heart rate chart:

> *It shows straightforward my exercise intensity, giving me a feeling of what is a high intensity workout…I did a great workout session on that day, very intense. I looked at it (the heart rate record) after going back home. I knew by this way if I gave my all or held something back…if heart rate is in the right range, I have to bite my teeth and fight to the end.*

*Although being busy I couldn't exercise that much, the heart rate helped me to maintain*

*my performance, even when I don't make progress, I can make sure it stays the same.*

Jun, also in his early 30s, has a different approach. He worked as a software engineer for a tech giant known for its notorious working schedules but also high chances of promotion and cash bonus after a "good year." In exchange of such lucrative payback, Jun endured long working hours and performance pressure from his manipulative boss. One day, during his argument with his boss, his smart watch vibrated – abnormal heartrate. He said, although he felt stressed and burnout pretty frequently, the warning turned out to be the final straw. He quit that job a few weeks after our first interview.

Heart rate is a good example of multiplicity of practices, and how different knowledge, attitudes, and skillsets are associated with the same data. For some users, heart rate provides a relatively convenient and straightforward marking (among many other metrics, standards, and plans that users might find difficult to follow) that reflects the "work" that users have done. Watches and wristbands categorize certain heart rate zones as "cardio training" or "fat burning." Users gauge their intensity of physical activities with heart rate to fulfill such purposes: burning calories or training cardio. Some users talked through their fitness routines using screen captures of heart rate diagrams that apps generate after each day or workouts. They recollected their own performances with a sense of achievement, that one pushes their body into doing something challenging, and that the body is capable of it or progressing towards it. Some users referred to their long-term physical status and goals regularly to see if their bodily capacities are improving. Yue, in his 30s and working for a tech giant in Hangzhou, has been monitoring resting heart rate for a few years. He scrolled through his 3 years of data, showing that he was currently working on returning the number to around 60 which he maintained a few years ago while being active.

Rather than subjecting themselves to overall and seamless self-surveillance that requires extensive labor, those performance-seeking users make the same set of practices ritually and habitually during designated workout sessions. To maintain the progress, users tend to develop a stable habit to concentrate their bodily work on and around fitness activities and/or diet plans. Data are produced within certain boundaries and body is treated as a working project. The process of reproduction (such as resting and eating) is also translated by bodily metrics as something to consciously and effectively work on (such as Oura Ring's "readiness score," a metric/algorithm that indicates the need for rest). The data-knowledge production process – what does it mean by/how to achieve ideal performance metrics – is also conducted in designated spaces, such as the gyms classes or training sessions. Within the boundaries, the performance metrics are treated as a matter of fact. Based on the meanings and materiality of performance data, relevant skills and competencies are cultivated – the ability to filter, read, and understand relevant data, as well as the routines to keep track and monitor progress in a long term to gain rewards and positive feedbacks.

> *I care more about the performance, it (the Watch) keeps records of pace and heart rates.*
> *If there are improvements after a period of time, I will feel rewarded by gaining proof of*
> *my ability. (#2-30, male, in his 30s)*

> *I will pay attention to the long term, such as the weight records on Keep; and the*
> *persistence brings me improvement in performance and achieve a higher level. (#2-24,*
> *male, in his 30s)*

There is another mode of practices around the same data. Users describe their data along with their bodily forms, sensations, and living experiences, as something they'd like to keep and maintain. It is shown frequently in an overworking context that the aim of using data is to fight against fatigue and stress, and to find a healthy fix for a vulnerable body in which the productivity-oriented metrics are less important than the actual experience of vitality.

For example, the maintenance and restoration of health are, in many cases, a descriptive "status-being" (*zhuangtai*, 状态) – that one can keep a stable energy level and physical form along with a good mentality in a stressful workplace. Some people referred this word to the living experiences that connect data to bodily sensations rather than achievements. The heart rate numbers are used to describe or correspond with their physical and mental health (or lack thereof) in and around their physical and social environment.

> *The first thing is to check my sleeping data and (resting) heart rate. If it looks great, that means I have a relatively good status-being...to support my day of hard work. I won't blame myself for it (not resting enough), but maybe I'll slack off when it's (the number) bad. (#2-49, male, in his 30s, works for a tech giant)*

> *I noticed that recently that if I work overtime to 10 p.m., my heart rate starts to dwindle, and I would feel bad. It is a reminder for me to get up, get around, and take a breath. (#2-50, female, in her 20s, works for a planning agency)*

The relationship between subjective senses (tiredness, fatigue, or rested) and the material world is addressed and interpreted using some objective data. It is collected through daily routines rather than from designated exercises. There is hardly any boundary, spatial or temporal, around where personal data is collected, interpreted, and enacted; while the meanings are personal and act as a reference to their work/life situations (such as a sedentary lifestyle due to the lack of mental and physical resources) rather than a disembodied fact which needs more "work."

The enactment of the data, aiming at maintaining a sound status-being amid stressful overwork, is also different from pursuing bodily performance. The data practice usually lacks regularity, is scattered and infrequent. Some interviewees mentioned that chasing after performance metrics gave them anxieties. The skills are different: They usually "take a look" at the numbers and "keep in mind" of their status being, seeking sense of reassurance if numbers don't fluctuate or go bad too much. It is important not only to comprehend "scientific" data as fact and work towards goals, but also to remind themselves what situations they are in, what strategies they are taking, and what kind of maintenance their body needed.

The description of such status-being shares similarities with the traditional Chinese body-mind relationship, depicting a rather holistic picture of a tired, vulnerable body burdened by overworking

culture – both from work and excessive self-care. The quest for care translates into languages of knowing and balancing rather than engagement and improvement; sometimes, the data is enacted with a sense of resistance against overworking. One interviewee, a software engineer at a big tech company, said about the "Body Battery," an algorithm developed by Garmin smart watch:

> *I look at the number to remind myself that I am tired. In this culture (of overworking), people feel reluctant to admit their tiredness, they need to work harder and harder so that they deserve something. But the number is like a valid proof that I also deserve rest.*

Users also talked about looking back at the accumulated data to summarize their past living experience and its connection with past or current health. An interviewee (#2-79) talked about moving from Hangzhou to Beijing and how his environment has changed with data from Xiaomi wristbands:

> *you can see the running and walking tracks when I was in Hangzhou; those are all around where I stayed…I was rather active, and my body weight was kept well. But after moving to Beijing, there is nowhere to go for a walk and run.*

Users hold doubt about the accuracy of the bodily data, the validity of the algorithms, yet sometimes pick a couple of metrics to "see the trends" and "swiftly sense the changes" (#2-32) while keep flexibility of their schedules as well as interpretations to avoid anxiety or disappointment. Some of the users treat data as a tool to build regularities of their routine, or to grasp a brief idea of their bodily status, then to reduce the frequencies of data entries or readings (rather than to increase or habituate the technological interactions as many products and services would encourage). A busy physician talked about his data entries of calories, both intake and consumption:

> *My baseline is that the consumption is more than my daily intake, or doesn't exceed too much. After a period of training and recording, I kind of have a "muscle memory" of it (and don't have to do it every day). (#2-42)*

It is not the picture that technology makers would envision – a responsible, knowledgeable user that aligns their actions to the intentions of growth; but a user who deploys a set of skills, with partial, incomplete or even inaccurate data to recognize their vulnerabilities, to connect and confirm bodily sensations, and to improvise solutions and strategies accordingly. Most importantly, users' data and interpretations of bodies are co-created as the interface through which users interact with their environmental and social context, drawing a holistic picture of their embodied living experiences with the language of "status-being."

*Table 2 two modes of data practices*

| Practices | Data as "performance" | Data as "maintenance" |
| --- | --- | --- |
| **Data is read as** | Goals, plans, performance gauge | Bodily "status-being", life reviews |
| **Data collected through** | Designated activities | Mundane routines |
| **Data interpretation** | Scientific, factual, objective | Personal, referential |
| **Technological interactions** | Frequent, habituated | Scattered, infrequent |
| **Body as** | Working subject | Needing protection |
| **Selfhood is built through** | Gaining control of work and life | Balancing stress |

The two modes of practices are not fixed nor exclusive. Users oscillate between the two, or even combining different elements from the two to create their own narratives, showing *multiplicity* rather than differences. These two modes correspond with how they view the relationship between overwork and selfhood – as X. Li (2022) has shown, those who work in "996" companies will overwork as a way of cultivating a "self" by gaining a sense of ownership over their bodies (performer); or see the inevitable overwork as trading their bodies for high salary or status, thus necessary to maintain a sustainable and sound status-being to avoid burnout (maintainer). The multiplicity of performing and enacting data practices confirms the dialectic relationship between work, body, and self.

Although performers tend to interpret the data into achievements and accomplishments, it is not to say that the "performance" practice is an extension from the culture of productivity. The building of bodily performance is context-dependent and embedded in the self-work relationship. Notably, users separate their fitness activities from work in and around workplaces, creating an exclusive, controllable environment that they can choose to make sense of certain numbers. For some people, the data represents not the result, but the process of bodily work – a way of gaining control of their bodies by performing metrics. Several avid fitness goers working overtime talked about their incentives of working out – to create a disciplined, controllable body against uncertainties of precarious work, and to find something outside of work to accomplish. Body as the "only thing that can be controlled," (#1-13) and the accomplishment from bodily work is objective, concrete, and owned by themselves. The stressful nature of work needs a place to detach mentally and physically, and the bodily work and "achieving goals" increases the sense of self-efficacy, "having more confidence in doing other things" (#2-5).

Those who spoke more about maintenance were more explicit about resistance against "involution." They are aware that the overworking is unavoidable, yet "trading health for money" is unsustainable. Users expressed the necessity of shielding the vulnerable body against anxiety, stress, and overworking culture by calibrating a status-being that is moderately active, resilient, and sustainable. It helps them exit from the vicious cycle of "involution," both at work and of self-care. An interviewee (#2-37, female, in her 30s, just quit a stressful digital media job) described a never-ending cycle she was in – overwork gets her a better pay, but brings stress and overeating, which lead to excessive spendings on fitness service and healthy meal package. "I hoped to earn more, so I worked harder. But stress (from work) had me buying health instead." Now she has abandoned the pursuit to improve her data, but pays more attention to how life has changed that brings positive changes in data.

In sum, the difference between these two practices is how control and autonomy are perceived under current social and political economy, and what can be done to fix, or resist, the never ending "involution" of productivity and responsibilities. Building a body of higher performance and capacity is to believe in the agential capacities (Lupton & Smith, 2018) of working, hacking, and fixing problems according to metrics and data. It is self-initiative, but also serves as an alternative of work/productivity-centered self, a self that is partially free from organizational control. Users on the other side, in contrast, view control and autonomy as the ability to resist "involution" and to find a balance despite the unavoidable fate of overwork. The attitude of "lying-flat" (*tangping*, 躺平, means giving up), both in work and workout, is not a moral judgement of being lazy or irresponsible. It is a contextual recognition of the environment and the ability and necessity to think and act – just as Jun did when he talked about his abnormal heartbeat. It was certainly not his heartbeat (or health risks, as Apple Health might suggest) that lead to his resign, but the recognition of the toxic overworking environment around him. Health is interpreted not only as the physical or mental wellbeing of the body, but also situational status that someone strives to find control and navigate through the era of "involution," COVID restrictions, and economic downturns.

**Discussion and Conclusion**

In the Chinese tech sector, the fast-paced competition has rewarded the young and engrossed yet leaves many unhinged and to the detriment of their health. In a culture of overwork, the need for self-actualization creates performance anxieties and never-ending cycle of "involution," in which the challenge of self-care has been translated into language of productivity. One must invest extra, strenuous effort to guarantee a healthy body that is capable of handling stress and navigate overwork. Personal data collected throughout the series of hard bodily work is interpreted as personal achievements and accomplishment, and users have cultivated various skills and habits in a demanding environment to secure the production and reproduction of bodily value. In a neoliberal economy, the ethics and responsibilities of health are concentrated onto health promotions, personal choices, and the values and optimization of bodies under a biomedical gaze (Carter et al., 2018; A. E. Clarke et al., 2003; Lupton, 1995; Rabinow & Rose, 2006). The manifestation of productivity in self-tracking in China, as well as the origins of "involution" both in workplaces and personal lives, reveals another layer of technological

practices – the making of an authority-approved, market-savvy, "civil" citizen who takes the pursuit as not only a responsibility, but also a privilege.

However, the strenuous work and "involution" conflicts with limited mental and physical resources. Therefore, users apply a certain level of interpretive flexibility around the enactment of personal data. Some of them anchor their bodies to performance metrics that are part of a capable body and self, gaining control of their life amid uncertainties and organizational control. Some others choose to treat data as a reflection of their physical and mental status-being and calibrate a balance between work and life – or in some cases, guarding their life and vulnerable body against overwork and "involution" of growth and competition. This paper thus argues against that self-tracking is merely an instantiation of neoliberal self-surveillance. It carries certain elements of neoliberal governmentality which centers on the making of selfhood, yet the technological practices of self-tracking serve as an interface of socio-political conflicts in the personal life-world (Habermas, 1984). The enactment of personal data mediates the relationship between individuals and their social and physical environments, especially during conflicts and contradictions. Human cognition and emotion thus get embedded in this process as certain metrics collected and read not only by digital sensors, but also through embodied actions and interpretations that interact with the social context. Such context is, therefore, not to be ignored or universalized, but need to be a critical part of examination. Recent studies have focused on differences and agencies in self-tracking (Dolezal & Oikkonen, 2021; Sax, 2021) and taken consideration of situated objectivity and multiplicities (Pantzar & Ruckenstein, 2017; Pols et al., 2019). This paper connects personal data practices further with the underlying socio-political dynamics and power structures that channel personal living experiences and responses. The multiple practices reflect not only the difference, but how individuals interact with their social environment and situate themselves in the politics of everyday life.

People often argue around whether certain wearable or fitness tracking devices are "useful," but rarely dig deep into the meanings and modes of use. Health outcomes are central to personal and public health; but how the outcomes could be attained and sustained in certain social and environmental context remains a long-term concern. Technological terms such as "user retention" measure whether or not a user continue to use certain devices or applications, but ignore or flatten multiple modes of use into engagement numbers. Another term, "user compliance," implies a moral judgement that non-intended use of certain devices is a fault that needs to be corrected. The underlying power mechanisms need to be questioned – who is guiding the collection of data? Who defines the meaning of data? In line with these sociological critiques, this paper highlights the importance of situated data. Technological practices are deeply cultural and even political - data carries assumptions and meanings, and entangles with personal practices that carry out, appropriate, or sometimes reject certain meanings. It reveals the limitation of commercial technological imaginary, and the necessity for study of digital health among various groups of people who are situated in different cultural and socioeconomic context.

Future research around different demographic, race/ethnicity, and social classes and their unique practices around personal data and their navigation of socio-political structures will reveal more valuable insights on the entanglement of human and technologies. It is also necessary to reframe or re-problematize health practices with "*thick data*" – data that links to environment, social/cultural aspects, and user's agency. Rather than users' retention and compliance, studies should focus on how users form and change their mode of using and develop measurement to reflect multiple modes and contexts. Users' knowing and doing should be prioritized, and to guide social-sensitive designs that benefit diverse populations and empower users with their agencies and capacities.